# Integrated, ultra-compact high-Q silicon nitride microresonators for low-repetition-rate soliton microcombs

*Zhichao Ye, Fuchuan Lei, Krishna Twayana, Marcello Girardi, Peter A. Andrekson, and Victor Torres-Company\**

Department of Microtechnology and Nanoscience
Chalmers University of Technology
41296 Gothenburg, Sweden
E-mail: torresv@chalmers.se

Multiple applications of relevance in photonics, such as spectrally efficient coherent communications, microwave synthesis or the calibration of astronomical spectrographs, would benefit from soliton microcombs operating at repetition rates <50GHz. However, attaining soliton microcombs with low repetition rates using photonic integration technologies represents a formidable challenge. Expanding the cavity volume results in a drop of intracavity intensity that can only be offset by an encompassing rise in quality factor. In addition, reducing the footprint of the microresonator on a planar integrated circuit requires race-track designs that typically result into extra modal coupling losses and disruptions into the dispersion, preventing the generation of the dissipative single soliton state. Here, we report the generation of sub-50GHz soliton microcombs in dispersion-engineered silicon nitride microresonators. In contrast to other approaches, our devices feature an optimized racetrack design that minimizes the coupling to higher-order modes and reduces the footprint size by an order of magnitude to ~1mm$^2$. The statistical intrinsic Q reaches 19 million, and soliton microcombs at 20.5GHz and 14.0GHz repetition rates are successfully generated. Importantly, the fabrication process is entirely subtractive, meaning that the devices can be directly patterned on the Si$_3$N$_4$ film. This standard approach facilitates integration with further components and devices.

## 1. Introduction

Dissipative Kerr soliton (DKS) microcombs[1] with THz repetition rates can be readily obtained in lithographically patterned structures thanks to the small bending radii attainable with large-index-contrast structures. Such THz-rate microcombs can be very broad[2–4] – even span an octave.[5–7] DKS microcombs operating at smaller repetition rates, i.e. comparable to the electronics bandwidth,[8–11] also find their unique set of applications in high-spectral-efficiency optical transimission,[12–14] the synthesis of microwave signals,[9,15,16] or the readout of THz microcombs.[17] The calibration of astronomical spectrographs also relies on frequency combs with repetition rates ~10-30 GHz,[18,19] as this corresponds to a line spacing matched to the resolving power of the spectrograph. Notwithstanding, generating and maintaining low-repetition-rate DKS microcombs is extremely challenging. Since the intensity buildup in a microresonator is proportional to Q/V,[20] operating with large cavity volumes V requires microresonators with large quality factors (Q).

Soliton microcombs with low repetition rates have been achieved in whispering gallery mode resonators, which display extremely large quality factors[8,21,22] but their planar integration is very challenging.[23,24] In integrated platforms, the high index contrast between core and cladding materials results in an increased susceptibility to scattering losses arising from nanometer-level roughness, thus limiting the achievable Q values. Indeed, successful generation of < 50 GHz DKS integrated microcombs has only been reported in planar silicon nitride devices fabricated with the Damascene reflow process,[9] with statistical Qs about 23 million. In this fabrication technique,[25,26] the $SiO_2$ substrate is pre-patterned and reflowed prior to $Si_3N_4$ deposition. Thick $Si_3N_4$ waveguides with ultra-smooth sidewalls can be readily achieved, but this technique results in a nonuniform waveguide height across the wafer due to an aspect ratio dependent etching.[27,28] Alternatively, $Si_3N_4$ waveguides can be fabricated using traditional subtractive processing methods,[29] where the $Si_3N_4$ film is lithographically patterned. The aspect ratio dependent etching can be simply overcome by applying extra etching time during the dry etching process, thus facilitating the co-integration with other devices and components.[29] Dispersion-engineered silicon nitride microresonators with Qs exceeding ten million have been reported by different groups,[30–32] but demonstration of DKS microcombs operating at sub 50 GHz has remained elusive.

An additional crucial aspect of sub 50 GHz DKS microcombs is the dramatically increased footprint of the microresonator. An overview of low-repetition-rate DKS microcombs based on different material platforms and their device footprint are summarized in Figure 1. The footprint of a microring resonator increases with $1/FSR^2$, and for a 10 GHz soliton comb based on $Si_3N_4$, the real-state area exceeds 20 mm$^2$. Such a large footprint is one of the key drawbacks of photonics integration when co-integrating with electronic circuits.[33] Alternatively, a microring resonator can be wrapped to form a compact racetrack-shaped microresonator to reduce the footprint.[34] However, coupling from fundamental transverse mode to higher-order modes (HOMs) can be easily introduced by the discontinuity or abrupt change of the curvature along the waveguide length. The coupling results in avoided mode crossings which can prevent soliton formation.[35] In order to filter out the HOMs, a microresonator with tapering to single mode section was introduced.[36,37] Unfortunately, since a narrow waveguide width is required at the single-mode section, the intrinsic Q drops to below 10 million. In addition, the tapering modifies the dispersion along the length of the resonators, making it difficult to access the single soliton state.[37] In the field of photonic integration, there is a bulk of research that focuses on adiabatic bend designs in order to dramatically reduce bending radius, bending loss and coupling to HOMs.[38–40] However, up to date, it remains unknown if these adiabatic bend designs could suppress the coupling to HOMs to a sufficiently low level in high Q microresonators and allow accessing to DKS states.[41]

In this work, we designed "finger"- and "snail"-shaped microresonators to dramatically reduce the footprint by an order of magnitude, see Figure 1. By properly designing bends in microresonators, the coupling from fundamental mode to HOMs is significantly suppressed, and the statistical intrinsic Q reaches 19 million. Soliton microcombs with repetition rates below 50 GHz (down to 14 GHz) are achieved for the first time in $Si_3N_4$ microresoantors fabricated by a subtractive processing method. Preliminary results of this work were presented in [42]. Here we provide a more detailed study including radio frequency (RF) beatnote measurement and single-sideband (SSB) phase noise characterization of the photodetected repetition rate, and present more advanced results including higher $Q_i$ and more compact microresonator designs.

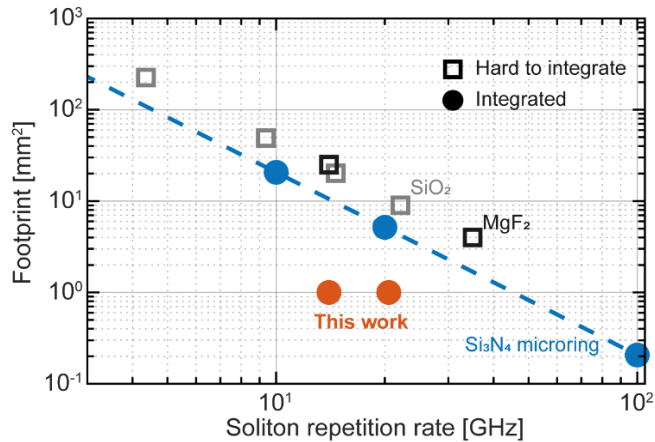

**Figure 1.** Overview of the repetition rate of DKS microcomb achieved in different platforms and their corresponding device footprint. The dashed blue line indicates how the footprint of $Si_3N_4$ microring resonator scales with the soliton repetition rate. The references for the comparison are [22] for $SiO_2$, [15,21] for $MgF_2$ and [9,43] for $Si_3N_4$.

## 2. Microresonators design, fabrication and characterization

We designed three racetrack-shaped microresonators: 1) a finger-shaped microresonator "A" using circular arc bends to connect straight waveguides; 2) a finger-shaped microresonator "B" using adiabatic bends to connect straight waveguides; 3) a snail-shaped microresonator "C" consisting of Archimedean spiral waveguides and adiabatic bends. The footprint of each microresonators is limited within 1 mm$^2$. The Si$_3$N$_4$ microresonators were fabricated using a subtractive process method described in [31]. The waveguide height and width are 740 nm and 1900 nm, respectively. An identical geometry was used for both bus waveguides and microresonator waveguides in order to achieve high coupling ideality.[44] Compared with [31], a multipass electron beam lithography (EBL) was introduced to further reduce the propagation loss,[45] and Si$_3$N$_4$ samples were annealed at 1200 °C in argon ambient to drive out the residual N-H bonds. The transmission spectra of the fabricated devices were obtained using frequency comb assisted diode laser spectroscopy.[46]

An optical microscopy image of microresonator A is depicted in Figure 2a. Here, circular arc bends connecting straight waveguides feature bending radii larger than 160 µm, i.e. large enough to neglect bending lossses. The measured integrated dispersion (defined as $D_{int} = \omega_\mu - \omega_0 - \mu D_1 = D_2\mu^2/2 + D_3\mu^3/6 + \ldots$ where $\omega_\mu$ is the angular frequency of the µ-th resonance relative to the reference resonance $\omega_0$, and $D_1/2\pi$ is the FSR) of the fundamental transverse-electric (TE$_{00}$) mode is shown in Figure 2b. As can be seen, mode crossings appear in most of the resonances, indicating severe coupling from the fundamental mode to HOMs. The coupling mainly results from the mode mismatch between straight and bent waveguides, even though the bent waveguides have very large bending radii. These avoided mode crossings not only introduce distortion in soliton spectrum but also potentially prohibit soliton comb generation.[35] These design issues are corrected in microresonator B, which also uses a finger shape, see Figure 2c. The adiabatic bends connecting straight waveguides are highlighted with red color. The adiabatic bends follow an algorithm to minimize the variation of the curvature, and consequently minimize the coupling to HOMs.[38] The curvature is defined as $k(s) = a_0 + a_1 \times s + a_2 \times s^2 + a_3 \times s^3$, where k is the curvature and s is the arc length. Boundary conditions based on the physical position, tangent, curvature and differential of the curvature at the start and end points are used to determine the polynomial coefficients. The minimum bending radius in the adiabatic bends was ~ 80 µm. The measured integrated dispersion of TE$_{00}$ mode is shown in Figure 2d. The retrived values from the fitting are $D_1/2\pi$ = 20.48 GHz and $D_2/2\pi$ = 30.0 kHz. The converted $\beta_2$ is -78.0 ± 0.2 ps$^2$/km. Compared with the integrated dispersion of microresonator A, eventhough bends with much smaller bending radii are introduced in microresonator B, the avoided mode crossings of microresonator B are significantly reduced, indicating that the adiabatic bends successfully minimize the coupling to HOMs. Although some weak mode crossings can still be observed, they are likely caused by random sidewall roughness introduced from electron beam lithography and reactive ion etching. Figure 2g shows the histogram of intrinsic linewidth of TE$_{00}$ mode family from 5 devices. The highest probable intrinsic linewidth ($\kappa_0$) is between 10 - 11 MHz, which corresponds to Q$_i$ of ~ 19 million. A representative resonance with intrinsic linewidth $\kappa_0$ = 9.1 MHz is shown in Figure 2h.

The microresonator C is snail-shaped, and consists of Archimedean spiral waveguides and adiabatic bends. Its optical microscopy image with adiabatic bends highlighted in red is shown in Figure 2e. The measured integrated dispersion is shown in Figure 2f. As can be seen, only weak mode crossings are observed, indicating the success of the microresonator design. The D$_1$/2$\pi$ and D$_2$/2$\pi$ values are 14.02 GHz and 13.4 kHz, respectively. The converted $\beta_2$ is -82.3 ± 0.1 ps$^2$/km. The footprint of this snail-shaped microresonator is only 1 mm$^2$. In fact, the proposed snail-shaped microresonator design is more universal than the finger-shaped microresonator since even smaller FSRs can be obtained by simply adding more cycles in the Archimedean spiral section. The transmission spectra of TE polarization of 5 devices were measured, and the histogram of intrinsic linewidth from TE$_{00}$ mode family is shown in Figure 2i. The most probable intrinsic linewidth is between 14 - 16 MHz, corresponding to a Q$_i$ of 13 million. The Q$_i$ is slightly lower than that of microresonator B since multipass EBL was not introduced in microresonator C. Nevertheless, the Q$_i$ is consistent with the value reported in Ref [31].

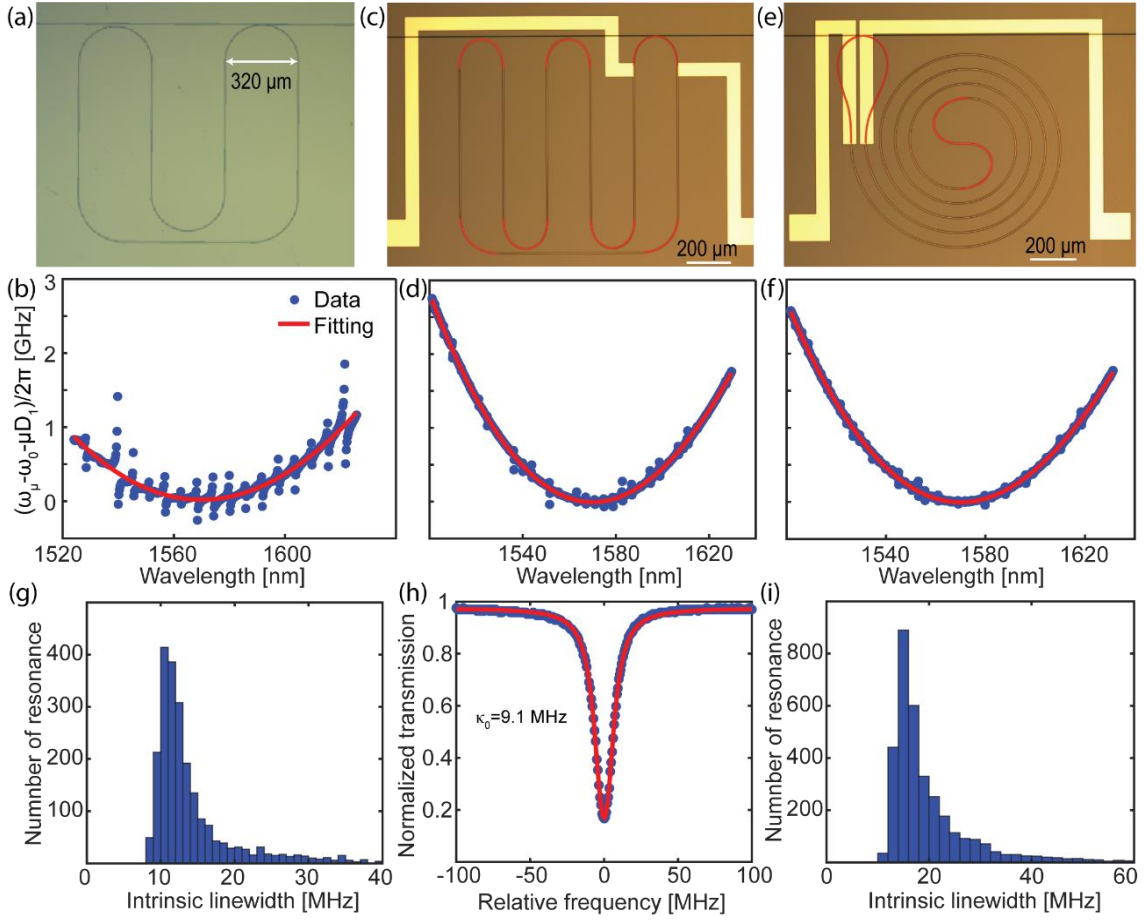

**Figure 2.** Microresonator design and characterization. a) Optical microscopy image of an incorrectly designed finger-shaped microresonator A, used as benchmark. The integrated dispersion of the $TE_{00}$ mode is shown in b). c), d) and e), f) idem to a), b) but for properly designed finger-shaped microresonator and snail-shaped microresonator respectively. g) Histogram of intrinsic linewidth from 5 properly designed finger-shaped microresonators B, and a representative resonance shown in h). i) The histogram of intrinsic linewidth values from 5 snail-shaped microresonators.

## 3. Low repetition rate soliton comb generation

We implemented soliton comb experiments based on the aforementioned three microresonators. While a soliton microcomb was unattainable from microresonator A, likely due to strong mode crossings,[35] DKS microcombs were successfully generated based on microresonators B and C. We pumped microresonator B at 1553.6 nm with an on-chip (off-chip) pump power of 224 mW (400 mW). The pump power was chosen to obtain relatively long soliton steps and wide soliton existence range, which significantly reduce the complexity of the experiment. The laser frequency was fixed, and a micro-heater was used to initialize the soliton comb.[47] The obtained single soliton spectrum with the pump wave suppressed by a fiber Bragg grating is shown in Figure 3a, with an inset figure showing the photodetected RF beat note. The repetition rate (20.5 GHz) of the soliton comb was directly measured by a high-speed electrical spectrum analyzer (ESA). A relatively strong Raman self-frequency shift of about ~ 11 nm was observed due to the small $D_2$ and large detuning between laser frequency and cavity resonance.[48] The snail-shaped micoresonator was pumped at 1553.6 nm with an on-chip pump power of 282 mW. A microheater was used to initialize the soliton comb, and the obtained single soliton spectrum is shown in Figure 3b. The repetition rate was measured to be 14.0 GHz.

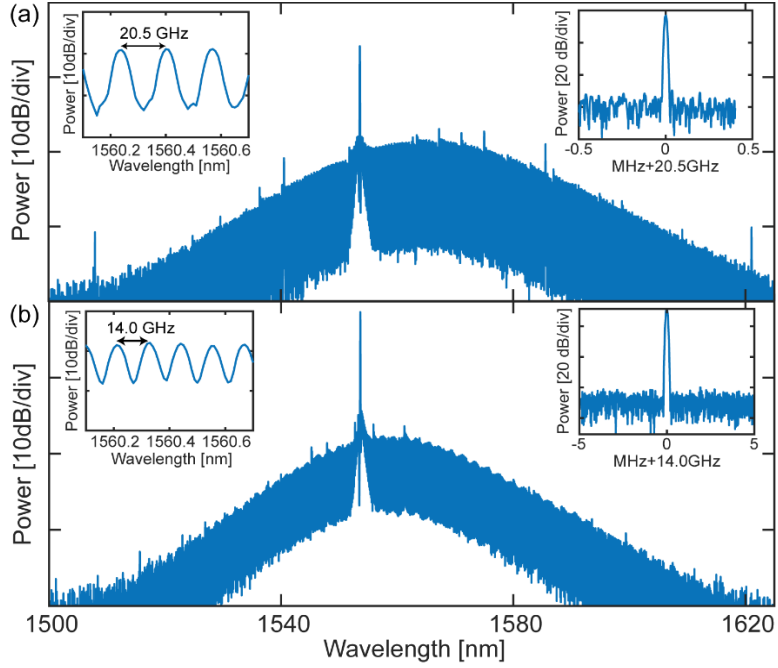

Figure 3. Single soliton microcombs with repetition rates of 20.5 GHz and 14.0 GHz. a) The optical spectrum of single soliton comb with a repetition rate of 20.5 GHz. The inset figures show a zoomed in optical spectrum and corresponding RF beat note measured with a resolution bandwidth of 1 kHz. b) idem to a) but for a single soliton with a repetition rate of 14.0 GHz.

## 4. Phase-noise characterization

In order to investigate how the stability of the soliton microcombs was affected by the racetrack-shaped microresonators, we characterized the phase noise of the photodected repetition rate microwave signal. We used the setup shown in Figure 4a. We only characterized the soliton comb with a repetition rate of 20.5 GHz since the soliton comb with 14.0 GHz repetition rate has strong pump power due to the limited suppression of the tunable fiber Bragg grating.

The frequency drift of tunable external cavity laser and the thermal drift of microresonator can result in undesired drift of detuning between the pump laser and cavity resonance, which consequently affects the repetition rate stability due to Raman self-frequency shift.[48] Therefore, we implemented a Pound-Drever-Hall (PDH) locking to control the detuning parameter. The PDH error signal was sent into a servo locking box, and the feedback signal actively controls the current of the pump laser. The single-sideband (SSB) phase noise of the microwave signal generated by the soliton microcomb was measured when the comb was operated at the detuning values 297, 323, 349, 375 and 399 MHz. The obtained SSB phase noise curves are shown in Figure 4b. The peak at 3 kHz is consistent with that observed in [9], and originates from the Toptica laser used in this experiment. A 'quiet point' [49] was found at detuning of 323 MHz (marked in red in Figure 4). When operating at the 'quiet point', the phase noise in the offset frequency range between 100 Hz to 10 kHz was significantly reduced. At offset frequency of 2 kHz, the phase noise was successfully suppressed from -65 dBc/Hz to -83 dBc/Hz. The SSB phase noise at offset frequency >200 kHz reaches a shot noise limited floor, which was estimated to be -119.5 dBc/Hz.[16] The obtained SSB phase noise is comparable to that obtained from $Si_3N_4$ microring resonators when the soliton comb was working at a quite point and PDH locking was implemented.[9] Purer microwave signals could be obtained by microwave injection locking and directly repetition rate locking[50,51] or with the use of higher power handling photodiodes.[52] The quality of the generated microwave signal indicates that the properly designed racetrack-shaped microresonator is feasible for stable soliton comb generation at sub-50 GHz while dramatically reducing the device footprint.

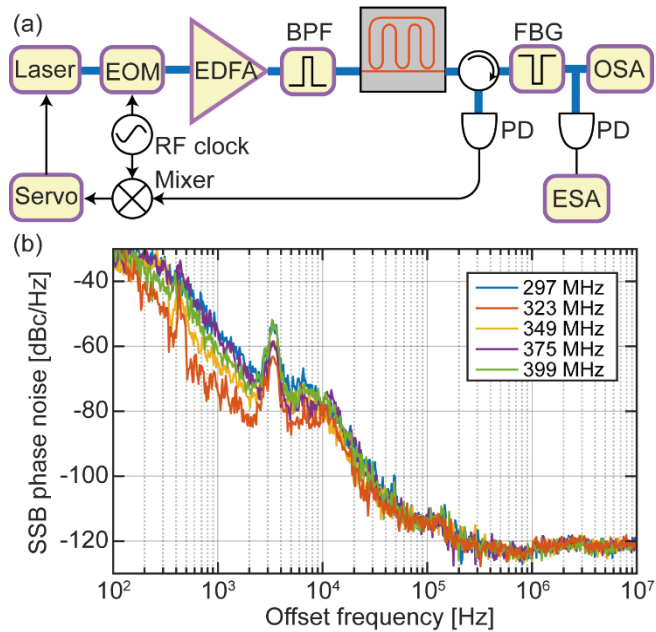

Figure 4. SSB phase-noise measurements of soliton microcombs. a) Experimental setup of phase noise measurement of soliton comb with PDH locking. EOM, electro-optic modulator; EDFA, erbium-doped fiber amplifier; BPF, optical bandpass filter; FBG, fiber Bragg grating;

OSA, optical spectrum analyzer; PD, photodiode; ESA, electrical spectrum analyzer. b) SSB phase noise of microwave signals generated by soliton combs with different values of detuning.

## 5. Conclusion

In summary, we have demonstrated low repetition rate (down to 14 GHz) soliton microcombs based on $Si_3N_4$ racetrack-shaped microresonators. These microresonators are fabricated using a more standard subtractive processing method, while statistical intrinsic Q reaches $19 \times 10^6$, sufficiently high to offset the drop of intracavity power when operating with large cavity volumes. With adiabatically designed microresonators, the coupling from fundamental mode to HOMs is significantly reduced, enabling the access to the single soliton regime. The footprint of the microresonator is an order of magnitude smaller than previously reported microring resonators with same FSR. Our compact microresonator design is compatible with the writing field (1 $\times$ 1 $mm^2$) of electron beam lithography which enables ~ 10 nm features that are unattainable by DUV stepper lithography. The repetition rate is sufficiently stable to enable applications in high spectral efficient coherent optical transmission, frequency synthesis and microwave photonics.

**Acknowledgements**

Z. Ye and F. Lei contributed equally to this work. The $Si_3N_4$ samples were fabricated at Myfab Chalmers. The authors acknowledge supports from The European Research Council (CoG 771410), the Swedish Research Council (2015-00535, 2016-03960, 2016-06077, 2020-00453) and the H2020 Marie Curie Innovative Training Network Microcomb (812818).

**Conflict of Interest**

The authors declare no conflicts of interest.